\pdfoutput=1
\documentclass[a4paper,10pt]{article}
\usepackage{pos}
\usepackage{graphicx}
\usepackage{subfig} 
\usepackage{float}
\usepackage{amsmath}
\usepackage{amssymb}
\usepackage[utf8]{inputenc}
\usepackage{url}
\usepackage{hyperref}

\DeclareUnicodeCharacter{2212}{-}

\title{String shoving effects on jets in p-p collisions}

\author*[a]{Smita Chakraborty}

\affiliation[a]{Department of Astronomy and Theoretical Physics, Lund University,\\
  Sölvegatan 14 A, Lund, Sweden}

\emailAdd{smita.chakraborty@thep.lu.se}

\abstract{

Di-jet observables are excellent probes to study the effect of jets in dense systems. Interacting Lund strings will affect jet observables and suggests a new common mechanism responsible for jet modification in p-A and A-A. In this proceeding, we present our new implementation of the string shoving mechanism in PYTHIA8 which lets us study the effects on jet observables in p-p and nuclear collisions. We also present preliminary results showing the effects in hadron-jet correlation studies.}

\FullConference{%
  HardProbes2020\\
  1-6 June 2020\\
  Austin, Texas}

\begin{document}
\maketitle

\section{Introduction}
The physical phenomena giving rise to collectivity effects in heavy ion collisions is still mostly attributed to hydrodynamic behaviour. For signals of flow in high-energy p-p system hence one cannot naturally extrapolate similar arguments, because of lack of corresponding inherent geometric substructure of the proton given it is a single nucleon. The answer to both these cases can lie in the bottom-up approach of the Lund string model, where evolution of the collision environment involves string degrees of freedom. 

In string shoving model in PYTHIA8, we use the string field between partons to calculate force between the string pieces. We argue that such a force will generate some flow in the system. Furthermore, such a force can alter jet observable to noticeable effect in heavy-ion collisions.

The feature of this method is a new frame of reference where geometrical bias of strings is eliminated, so that any pair of string pieces lie symmetrically with respect to each other. This simplifies the calculation of the interaction force and provides a base to study jets. We describe our new implementation of string shoving in PYTHIA8. We show preliminary results of two particle correlation in p-p and show effects of the interaction force on jets in p-p.




\section{String shoving in PYTHIA8}

The Angantyr model for heavy ion collisions in PYTHIA8 \cite{Bierlich:2018xfw} is based on default PYTHIA implementation for p-p, and Glauber based calculation for nuclear collisions\cite{PhysRev.100.242}, including Gribov corrections\cite{ALVIOLI2009225}. It gives good description of multiplicity and transverse momentum distributions in Pb-Pb collisions. It thus provides with the framework to further implement dedicated processes to account for additional heavy-ion signals like collectivity, and shoving is one such model being implemented.

Further work on to implement string interactions in Angantyr has been done in \cite{Bierlich:2017vhg}, where each string evolves into flux tubes that have been rendered with a Gaussian colour electric field $\mathcal{E}(r_\perp)$ inspired by lattice calculations\cite{Baker:2018mhw, Baker:2019gsi}, given by: 
\begin{equation}
\label{eq:Gaussian}
\mathcal{E}(r_\perp) = C \exp(-r^2_\perp/2R^2)
\end{equation}
where $r_\perp$ is the distance from the core of the flux tube, and R specifies its width and C is the normalisation constant. The interaction energy at a transverse separation $d_\perp$ between two flux tubes will be
\begin{equation}
    E_{int}(d_\perp) = \int d^2 r_\perp E(r_\perp) E(r_\perp − d_\perp) = 2 \kappa \exp( − d^2_\perp/4R^2)
\end{equation}
The force $f(d_\perp)$ per unit length is then 
\begin{equation}
\label{eq:force}
    f(d_\perp) = \frac{dE_{int}}{dd_\perp} = \frac{g\kappa d_\perp}{R^2} \exp \left( - \frac{d^2_\perp(t)}{4R^2} \right),
\end{equation}
where g is a tunable parameter with a value $\sim$ 1. The flux tubes are assumed to be parallel to each other, along the beam axis. 

 Note that two overlapping strings will give rise to mutual repulsion and not attraction. In an Abelian theory, we would have half the dipole pairs pushing each other almost half the time, and half pulling, depending on whether the dipoles are aligned or anti-aligned. A string segment is either a triplet(3) or an anti-triplet($\bar{3}$). Two parallel strings can be in a anti-triplet or sextet state, 3 $\otimes$ 3 = $\bar{3}$ + 6.  With anti-parallel strings in non-Abelian QCD($\bar{3} \otimes$ 3 = 1 + 8), we would only have string attraction in singlet cases, whereas the octet will give us a push. With parallel strings, on the other hand, we will have a push in the sextet case, and in anti-triplet case, it can give rise to one string via junctions. Which implies that in case of two dipoles with arbitrary colour charges on the ends, there are three cases: (a)1/18 of the time : strings pull (singlet), (b) 1/6 of the time : strings attract and form junctions, with no net effect to final state flow (anti-triplet) and (c) 7/9 of the time : strings push each other (4/9 from octet, 1/3 from sextet).
Correcting this small effect by getting rid of singlets with colour reconnection\cite{Bierlich:2015rha} will only be correcting a quite small effect. Thus 7/9 of the time overlapping strings push.

\subsection{Parallel frame formalism} 
After the hard interaction, when the strings evolve in space-time, each colour dipole formed between two massless  partons, are considered pairwise and are boosted from the lab frame to a new frame, where the strings lie in parallel planes and symmetric to each other \ref{fig:partons}. Such a Lorentz frame provides us with a symmetric topology to compute forces between a pair. The strings in this frame evolve in width to a maximum radius R($\sim$ 1 fm), from their production vertex. This evolution process continues until they reach the hadronization time or the maximum radius. 

In the parallel frame, each string has an opening angle $\theta$ that determines the length of the strings (separation between the two partons), and a skewness angle $\phi$, that is the off-set between the string pair such that they are not completely parallel (or anti-parallel in case of the reverse colour arrangement) to each other, but lie in parallel planes instead. The pseudorapidity $\eta$ and the skewness angle $\phi$ of the string-ends (partons) in the parallel frame can be expressed in the form:  
\begin{align}
	\cosh \eta = \frac{s_{13}}{4 p_{\perp1}p_{\perp3}} + \frac{s_{14}}{4 p_{\perp1} p_{\perp4}}, \\
	\cos \phi = \frac{s_{14}}{4 p_{\perp1} p_{\perp4}} - \frac{s_{13}}{4 p_{\perp1} p_{\perp3}} ,
	\end{align}
	where $s_{ij}$ are the squared masses for the partons $i$ and $j$, $p_{\perp i}$ is the transverse momentum of the parton $i$.
	
\subsection{Push distribution in hadrons} 
\begin{figure}[H]
    \centering
    \begin{minipage}[c]{0.32\linewidth}
    \centering
       \includegraphics[width=1.1\textwidth]{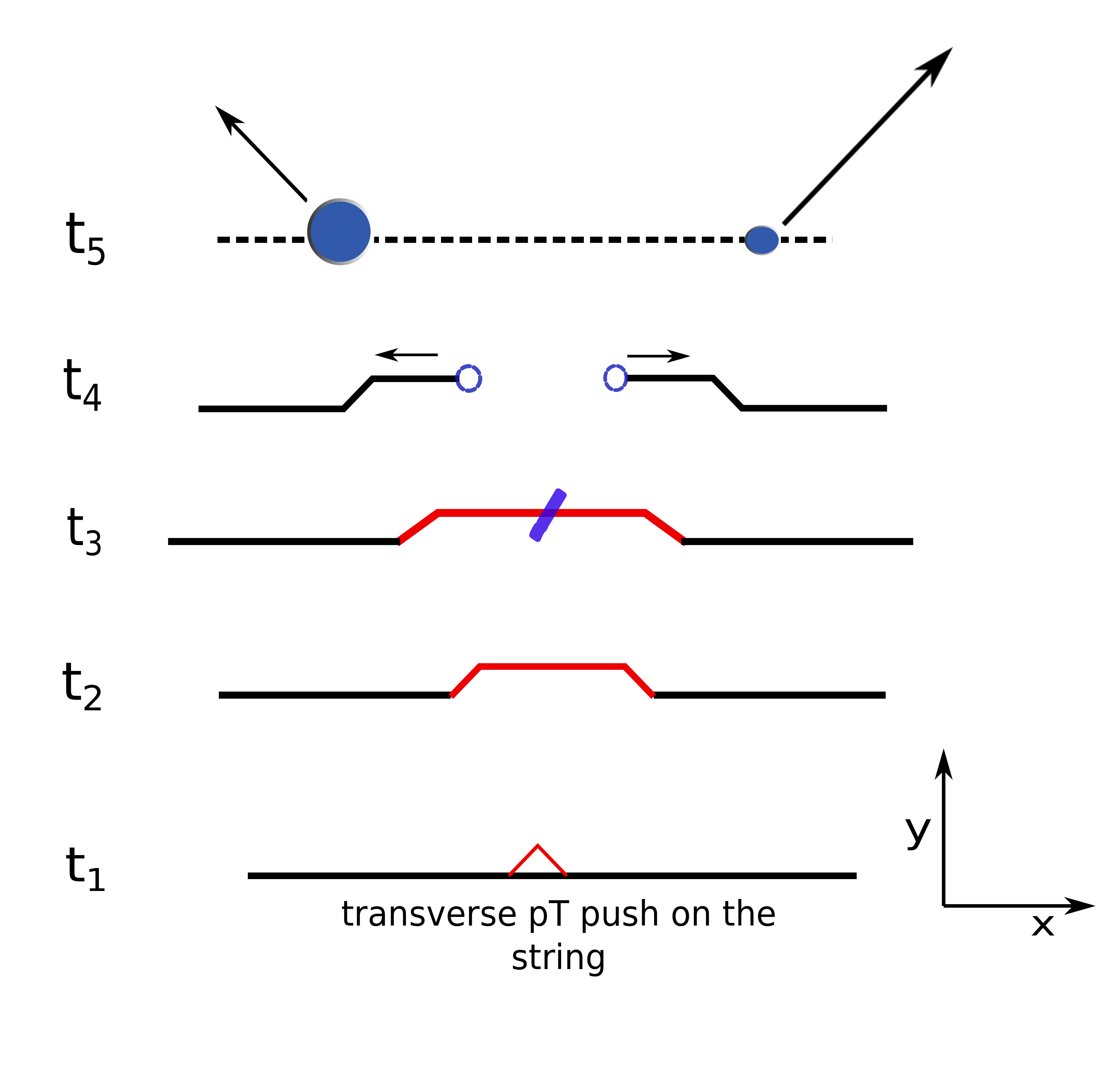}
    \end{minipage}  \begin{minipage}[c]{0.56\linewidth}
    \centering
    \includegraphics[width=1.2\textwidth]{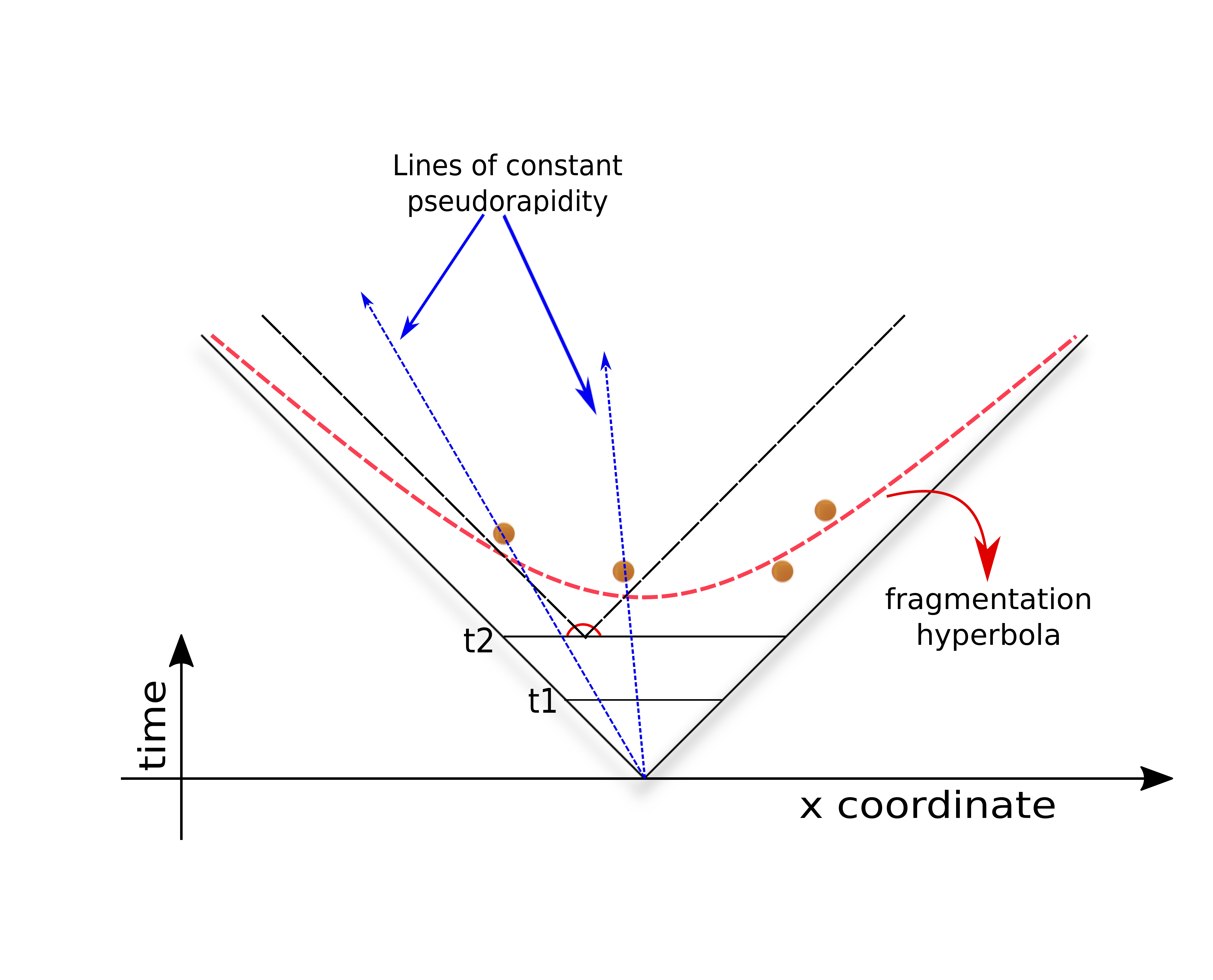}
    \end{minipage} 
    \caption{\label{fig:partons}Left: Schematic diagram of how a pT push spreads along the string piece in time steps, right: schematic diagram of a string evolution in x-t space, showing how a kink generated due to string interaction is divided as pT between two hadrons.}
    
\end{figure}
 While the strings evolve in radius, we want to veto the time steps leading to the final stage of string radius just before the string breaks and hadronizes. Vetoing the time evolution helps to calculate the force generated at each step and the string pieces are pushed after every iteration.
 Following the force given in \ref{eq:force}, the total transverse momentum push on the strings is: 
 $\Delta p_\perp =\int dt \int dx f(d_\perp(t))$.

We push the string pieces every time with a small fixed transverse momentum $\delta p_\perp$, following a probablity distribution $P(t)$, which will give the total push as:$\Delta p_\perp = \int dt P(t) \delta p_\perp$, in case $\delta p_\perp$ is small enough, we make the ansatz: $P(t) = \frac{1}{\delta p_\perp}\int dx f(d_\perp)$.

Figure \ref{fig:partons}(left) shows how a transverse push is distributed among all hadrons formed from the string regions where the push originated. Such a push exists as a small kink on the string piece, but as it has a  $\delta p_\perp$, it starts propagating along the string. The hadrons getting a share of the transverse push from this pT push will continue to move in their direction of initial pseudorapidity. Since this implementation is increasingly time consuming with increase in number of dipole pairs, the push is distributed after every $\mathcal{O}(10)$ such pushes are generated. 
\section{Results} 
\begin{figure}[H]
    \centering
    \subfloat{
    \includegraphics[width=0.5\textwidth]{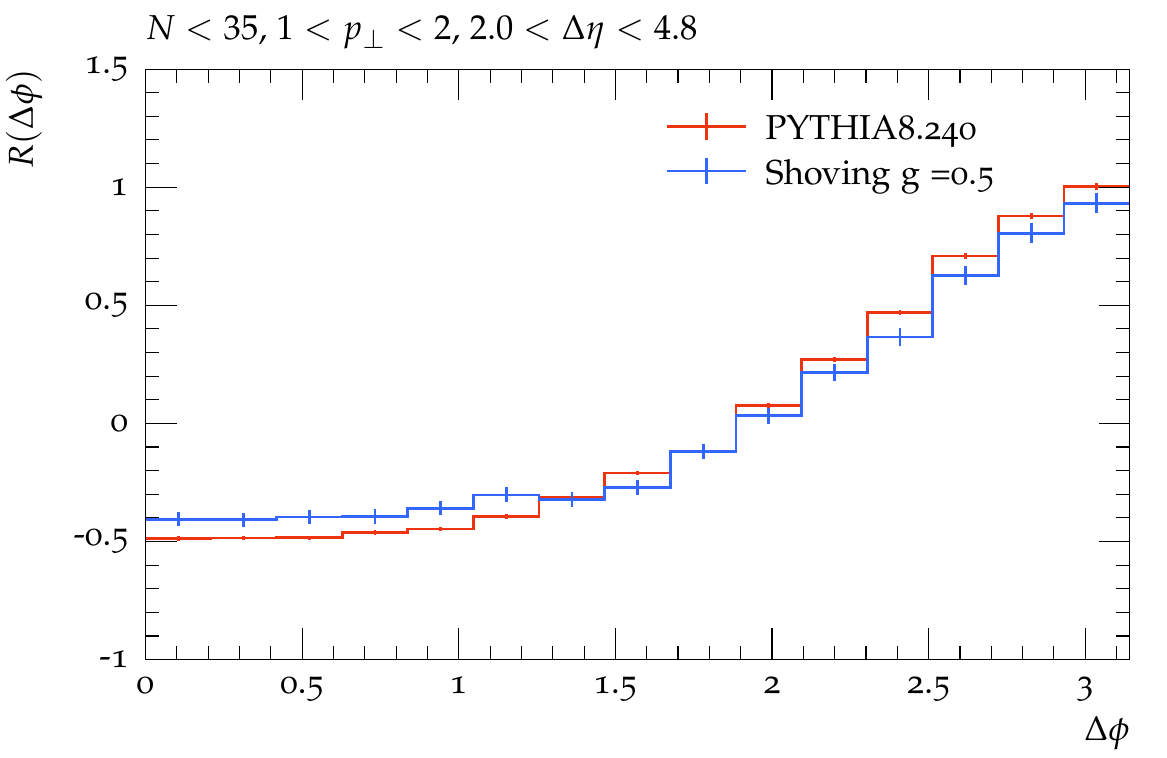}
    }
    \subfloat{
    \includegraphics[width=0.5\textwidth]{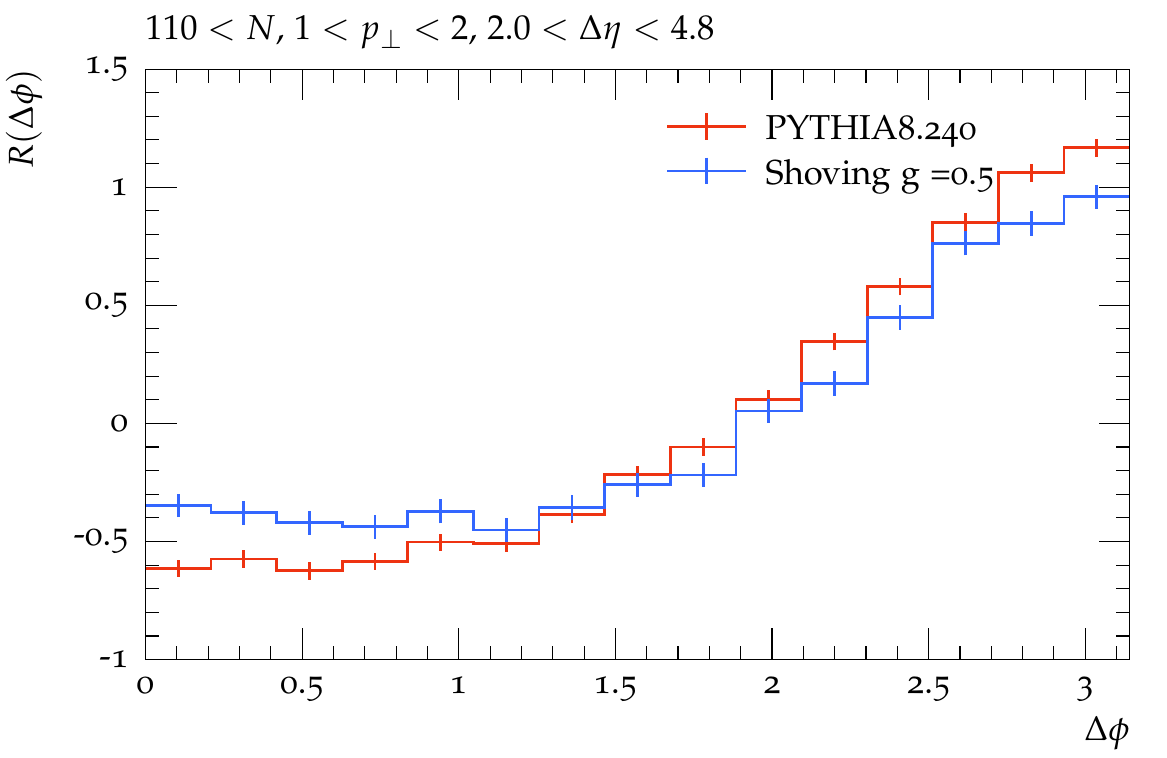}
    }
    \caption{Di-hadron correlations in p-p at 7 TeV at minbias for $N<35$ and $N>110$}
    \label{fig:correlation-plot}
\end{figure}

\begin{figure}[H]
   \centering
   \subfloat{
   \includegraphics[width=0.5\textwidth]{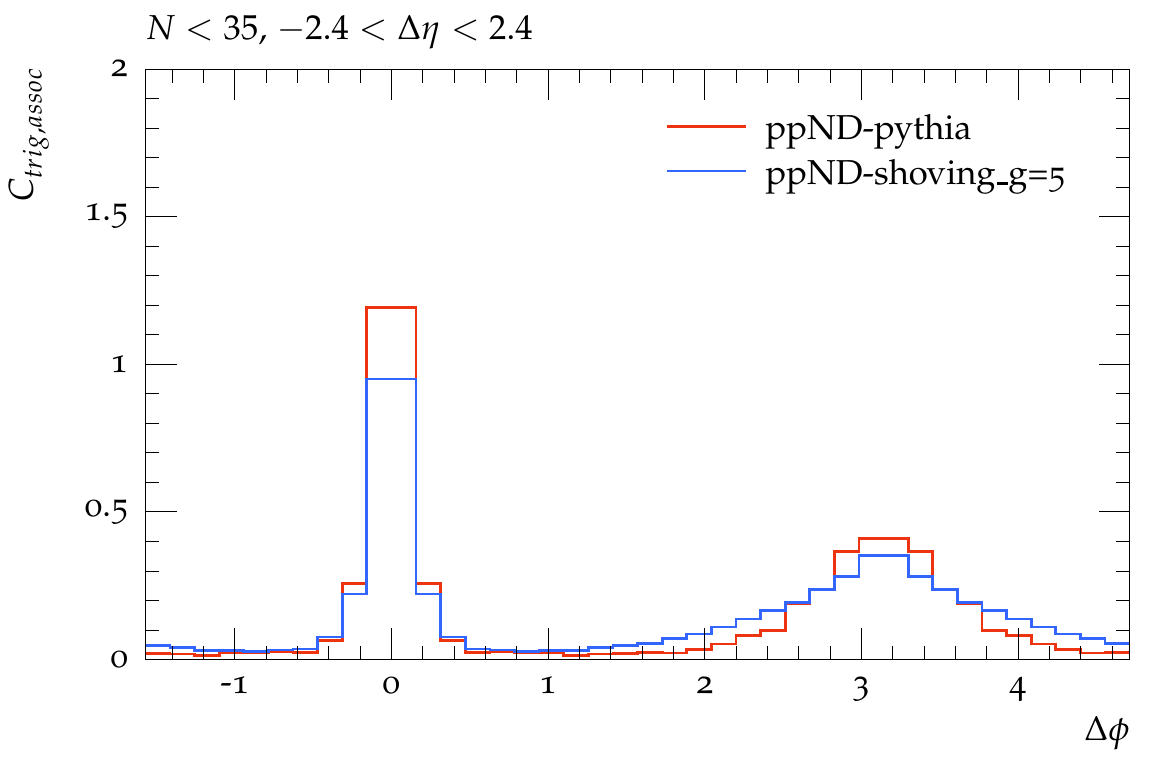}
   }
   \subfloat{
   \includegraphics[width=0.5\textwidth]{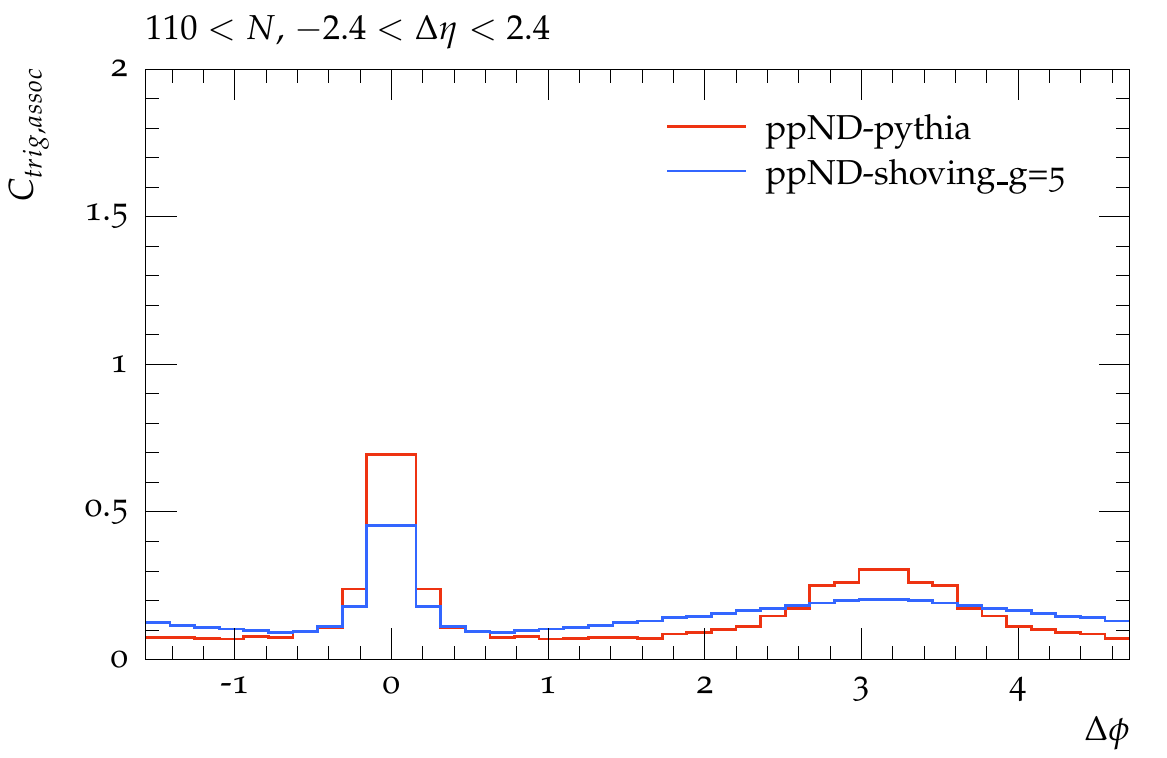}
   }
   \caption{ \label{fig:jet-plot} Charged hadron correlation in p-p at 7 TeV for g=5 for $N<35$ and $N>110$ for non-diffractive(ND) events }
  
\end{figure}

We present two sets of results, one which follows the study from \cite{Khachatryan:2010gv}(the p-p ridge) where we plot in figure \ref{fig:correlation-plot}, di-hadron correlation for proton-proton collisions for 1$<p_\perp<$2, 2.0 $<\Delta \eta<$ 4.8 for two multiplicity bins against difference in azimuthal angle $\Delta \phi$. The signal is constructed as $S_N$ = $\frac{1}{N(N-1)}$ $\frac{d^2 N^{signal}}{d \Delta \phi d \Delta \eta}$ and the background: $B_N$ = $\frac{1}{N^2}$ $\frac{d^2 N^{mixed}}{d \Delta \phi d \Delta \eta}$, the ratio of the signal to background is plotted R($\phi$) =  $\left< (\langle N \rangle -1)\left( \frac{S_N}{B_N} -1 \right) \right>$ where $\langle N \rangle$ is the number of tracks per event averaged over the multiplicity bin, and the final $R(\Delta\eta, \Delta \phi)$ is found by averaging over multiplicity bins. The result is compared to default PYTHIA8. We observe that the current shoving model reproduces the results as good as the earlier shoving implementation\cite{Bierlich:2017vhg}, where we see that p-p ridge has been produced as seen in data\cite{Khachatryan:2010gv}, which is more apparent in higher multiplicity like $N>110$.
The second set of results is to check if shoving can cause jet broadening and suppression, for that we increase the repulsion factor g to 5.0 to mimic the response in dense environments. We plot the azimuthal separation between a charged hadron picked between $| \Delta \eta | <$ 2.4 with 6 GeV $<p_\perp<$ 8 GeV and the associated charged hadron between$ |\Delta \eta|<$ 2.4 with 4 GeV $<p_\perp<$  6 GeV and compare to default PYTHIA8 for non-diffractive events. As we see in figure \ref{fig:jet-plot}, there is indeed a suppression in the near-side jet peak and a broadening at $\Delta \phi$ = $\pi$, the away-side jet peak. Hence the new implementation of shoving can give rise to jet modification in nucleus collisions and be used to analyze jet observables.
In conclusion, we observe that defining the parallel frame in string shoving model allows one to study events with jets. Furthermore, string shoving is a non-perturbative candidate that can give rise to collective effects and even cause jet modification in nuclear collisions.  
\section{Acknowledgement}
This work is done in collaboration with Christian Bierlich, G\"{o}sta Gustafson and Leif L\"{o}nnblad.
This work has received funding from the European Union's Horizon 2020 research and innovation programme as part of the Marie Skłodowska-Curie Innovative Training Network MCnetITN3 (grant agreement no. 722104).




\bibliographystyle{ieeetr}
\bibliography{HPproceeding}

\end{document}